\documentclass[%
 reprint,%
 aip,apl,%
]{revtex4-1}

\usepackage{bm}%
\usepackage[colorlinks=true,linkcolor=blue]{hyperref}%
\expandafter\ifx\csname package@font\endcsname\relax\else
 \expandafter\expandafter
 \expandafter\usepackage
 \expandafter\expandafter
 \expandafter{\csname package@font\endcsname}%
\fi
\usepackage{graphicx}
\usepackage{dcolumn}

\begin{document}

\preprint{APS/123-QED}

\title{Spin Hall angle in single-layer graphene}

\author{Juliana M. da Silva}
\affiliation{Departamento de F\'{\i}sica, Universidade Federal Rural de Pernambuco, 52171-900, Recife, PE, Brazil }
\affiliation{Departamento de F\'{\i}sica, Universidade Federal da Para\'iba, 58297-000 Jo\~ao Pessoa, Para\'iba, Brazil}

\author{Fernando A. F. Santana}
\affiliation{Departamento de F\'{\i}sica, Universidade Federal Rural de Pernambuco, 52171-900, Recife, PE, Brazil }

\author{Jorge G. G. S. Ramos}
\affiliation{Departamento de F\'{\i}sica, Universidade Federal da Para\'iba, 58297-000 Jo\~ao Pessoa, Para\'iba, Brazil}

\author{Anderson L. R. Barbosa}
\email{anderson.barbosa@ufrpe.br}
\affiliation{Departamento de F\'{\i}sica, Universidade Federal Rural de Pernambuco, 52171-900, Recife, PE, Brazil }

\begin{abstract}
We investigate the spin Hall effect in a single-layer graphene device with disorder and interface-induced spin-orbit coupling. Our graphene device is connected to four semi-infinite leads that are embedded in a {Landauer-B\"uttiker} setup for quantum transport. We show that the spin Hall angle of graphene devices exhibits mesoscopic fluctuations that are similar to metal devices. Furthermore, the product between the {maximum spin Hall angle deviation} and dimensionless longitudinal conductivity follows a universal relationship $\Theta_{sH} \times \sigma = 0.18$. Finally, we compare the universal relation with recent experimental data and numerically exact real-space simulations from the tight-binding model.
\end{abstract}

\maketitle

\section{Introduction}

The idea to manipulate information through spin was first introduced by Datta and Das \cite{dattaspintronica} and it has enhanced the development of spintronics in the last two decades \cite{PhysRevApplied.15.054004,jedema,lou,Hirohata_2014,Wang_2013}. 
This idea aims to use spin as an information carrier in place of charge, which makes high-speed computing possible.
Spintronics can be activated by spin-orbit coupling (SOC), which is the key to controlling spin transport properties without magnetic materials. SOC is a relativistic effect that is found in many branches of condensed matter physics.
The spin Hall effect (SHE) is one of the most significant phenomena observed in spintronics   \cite{spinhallh, perel, pereldois,doi:10.1142/S021797920603370X,RevModPhys.87.1213,Seifert_2018,Wang_2015}, which allows us to obtain a spin current from a charge current. 
More specifically, when a longitudinal charge current crosses a region with a strong SOC, it converts to a {transversal} spin current. 
Therefore,  the spin Hall angle (SHA) is an important parameter that is commonly used to quantify a material's ability to convert charge-to-spin currents. The SHA is defined as the ratio between the spin Hall current and the charge current, and its experimental values can vary from $0.01\%$ to $58\%$ for different materials in a disordered regime \cite{ando, hiroyasu,sagasta, fritz, pai, lou2, okano, zhu, wang, rezende, obed, zhang}.

The presence of disorder leads to universal spin Hall current fluctuations \cite{PhysRevB.72.075361,refId0,PhysRevLett.97.066603,PhysRevLett.98.196601,PhysRevLett.101.016804,PhysRevB.86.235112,PhysRevB.93.115120}, similar to universal charge current fluctuations \cite{RevModPhys.69.731}.
Therefore, it is necessary to investigate the SHA fluctuations to obtain information about the universality of converting a charge current into a spin current.
Ref. \onlinecite{PhysRevB.102.041107} started this investigation in an analytical, numerical and experimental analysis of metals. 
The authors show that for a quasi-unidimensional sample, the maximum SHA deviation $\Theta_{sH}$ follows a relationship with {dimensionless longitudinal conductivity $\sigma =N l_e/L$, where $N$, $L$ and $l_e$ are number of propagating wave modes, device longitudinal length and free electron path, respectively}, which is given by
$\Theta_{sH} \times \sigma = 0.18$.
{This proves that vanishing SHE---that is, when spin Hall conductivity is zero for any nonvanishing disorder strength \cite{PhysRevB.70.041303,PhysRevB.71.033311,PhysRevB.71.245327,PhysRevLett.96.056602,PhysRevB.72.075361,refId0,PhysRevLett.119.246801}}---is irrelevant for a realistic finite-size device where self-averaging over an infinite system size is avoided.

The use of two-dimensional materials, such as graphene, in electronic systems has been of great interest because of their ability to transport spin over long distances at room temperature, which optimises transport \cite{novoselov, risegraphene,RevModPhys.81.109, coloquiumspintronics,ingla}. 
However, it is well-known that graphene has a low SOC \cite{RevModPhys.81.109}. 
Therefore, several techniques have been proposed to improve its spin current generation capacity to circumvent this problem \cite{coloquiumspintronics,ingla,C7CS00864C,desordema,disorder,ISLAM2016304,10.1038/nphys2576}.

Ref. \onlinecite{10.1038/nphys2576} reported the first experimental realisation of the SHE in graphene. To increase the SOC, the authors added covalently bonded hydrogen atoms to graphene in a controlled manner. As a result, they obtained a SHE that is an order of magnitude greater than that observed in metals. This method also allowed measurements to be performed at room temperature, getting a value for the SHA at the charge-neutrality (Dirac) point (CNP) of $\Theta_{sH}= 58\%$. They then extended their studies to single-layer graphene doped with metallic atoms \cite{10.1038/ncomms5748}. The value obtained for SHA at room temperature was $\Theta_{sH}= 20\%$. This result shows that the interaction with metals produces equally strong effects, as observed in the former experiment. These experiments represent a significant advance in applying graphene in spintronics and stimulated other works to improve the SOC in graphene \cite{PhysRevResearch.2.013286, PhysRevResearch.2.033204,doi:10.1021/acs.nanolett.8b04368,WU2020396}.

Given the potential of graphene and the importance of studying SHA, the following questions arise: How do SHA fluctuations behave in graphene? And, what information do these fluctuations provide? This work shows that the relationship $\Theta_{sH} \times \sigma = 0.18$ is also valid for  single-layer graphene with disorder at the CNP, or any other Dirac material.
Note that concluding that this relationship is valid for metals and graphene is not straightforward because the mechanisms of charge-to-spin current conversion are different in both devices \cite{RevModPhys.87.1213,coloquiumspintronics}. 
To confirm our statement, we developed numerically exact real-space simulations from the tight-binding model with Bychkov-Rashba SOC, which explicitly violates $\vec{z} \rightarrow {-}  \vec{z}$ symmetry \cite{PhysRevLett.119.246801,PhysRevB.93.155104,PhysRevB.98.045407,PhysRevB.103.L081111}. Furthermore, motivated by Refs. \onlinecite{PhysRevB.77.193302, PhysRevB.82.121310,https://doi.org/10.1002/andp.201100253,PhysRevB.97.245307,SEIBOLD201763,KUMARSHARMA2021167711,PhysRevB.83.085306}, we analysed the behavior of SHA in single-layer graphene in the presence of uniform and random SOC. 
Finally, our analytical result is confronted with recent experimental data that can be found in the literature \cite{10.1038/nphys2576,10.1038/ncomms5748,PhysRevResearch.2.013286, PhysRevResearch.2.033204,doi:10.1021/acs.nanolett.8b04368}.

\begin{figure}
\centering
\includegraphics[scale = 0.3]{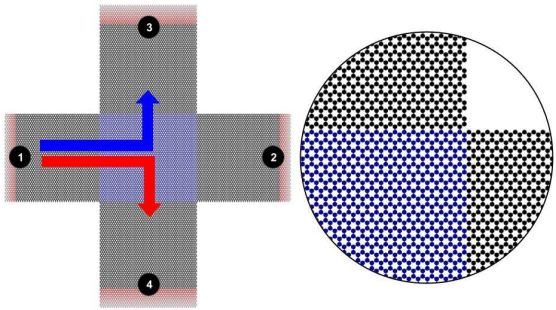}
\caption{The spin Hall device design of single-layer graphene. The scattering sample with disorder and strong SOI (blue) is connected to four semi-infinite leads.}\label{sample}
\end{figure}

This work is organised as follows. Section II gives the analytical result of the SHA fluctuations. The SHE in graphene is studied in the scope of the {Landauer-B\"uttiker}  model, which enables us to obtain the universal spin Hall current fluctuations in the CNP. In Section III, we compare the experimental data of SHA found in the literature with our result. In Section IV, we develop the numerical calculations to validate the analytical hypothesis that was presented in Section II. As shown in Fig. \ref{sample}, the SHE is numerically simulated in a honeycomb lattice sample (blue) that is connected to four terminals. Numerical calculations were implemented using KWANT software \cite{Groth_2014}. Finally, our conclusions are presented in Section V.

\section{Spin Hall angle} 

This section will describe our analysis of the spin Hall current and SHA. We will show that the SHA fluctuations lead to a universal relationship between maximum SHA deviation and dimensionless conductivity. Therefore, we divide this section into two subsections. In the first subsection, we introduce the SHE for a graphene device with disorder connected to four semi-infinite leads that are embedded in the {Landauer-B\"uttiker} setup for quantum transport and we then calculate the spin Hall current fluctuation at the CNP. In the second subsection, we develop an analytical calculation of SHA fluctuations.

\subsection{Spin Hall current fluctuations}

We designed a spin Hall device with four semi-infinite leads (black) connected to a scattering region with disorder and strong SOI (blue), as shown in  Fig.(\ref{sample}). 
The {Landauer-B\"uttiker} model can describe the SHE \cite{PhysRevB.72.075361,refId0,PhysRevLett.98.196601}. Therefore, the spin-resolved current through the $i$th electrode is
\begin{eqnarray}
I_{i,\alpha} = \frac{e^2}{h}\sum_{j=1}^4 \tau_{ij}^{\alpha} \left(V_i - V_j \right).\label{LB}
\end{eqnarray}
Applying an electric potential difference $V$ between leads 1 and 2, $V_1=V/2$ and $V_2=-V/2$, leads to a pure longitudinal charge current; see  Fig.(\ref{sample}). The spin-up carriers are then deflated to one side of the scattering region, while the spin-down carriers are deflated to the other. This leads to a pure spin Hall current in the transversal direction, see Fig.(\ref{sample}). The transmission coefficients $\tau^\alpha_{ij}$ can be obtained from  transmission and reflection blocks of the corresponding device scattering  $\mathcal{S}$-matrix, as follows
\begin{eqnarray}
\tau_{ij}^{\alpha} =\textbf{Tr}\left[\left(\mathcal{S}_{ij}\right)^{\dagger}\sigma^\alpha\mathcal{S}_{ij}\right], \quad 
\mathcal{S}=
\left[\begin{array}{cccc}
r_{11}&t_{12}&t_{13} &t_{14}\\
t_{21}&  r_{22}&t_{23}&t_{24}\\
t_{31}&  t_{32}&r_{33}&t_{34}\\
t_{41}&  t_{42}&t_{43}&r_{44}
\end{array}\right],\label{M}
\end{eqnarray}
where $\sigma^0$ and $\sigma^\alpha$ denote the identity and Pauli matrices, respectively, and with polarisation direction $\alpha = {x,y,z}$.

Following Refs. \onlinecite{PhysRevB.72.075361,refId0,PhysRevLett.98.196601}, we assume that the charge current vanishes in the transverse leads, $I^c_{i,0}=I^{\uparrow}_i+I^{\downarrow}_i$ for $i=3,4$ and that the charge current is conserved $I^c_1=-I^c_2=I^c$. From Eq. (\ref{LB}), we obtain a general expression for the transversal spin Hall current
\begin{eqnarray}
I^{s}_{i,\alpha} = \frac{e^2}{h}\left[\left(\tau^\alpha_{i2}-\tau^\alpha_{i1}\right)\frac{V}{2}
- \tau^\alpha_{i3}V_3 + \tau^\alpha_{i4}V_4\right],\; i={3,4},
\label{Is}
\end{eqnarray}
where  $I^s_{i,\alpha}=I^{\uparrow}_i-I^{\downarrow}_i$, and also for longitudinal charge current
\begin{eqnarray}
I^c &=& \frac{e^2}{h}\left[\left(4N+\tau^0_{12}+\tau^0_{21}-\tau^0_{11}-\tau^0_{22}\right)\frac{V}{4}\right.\nonumber\\&+& \left.\left(\tau^0_{23}-\tau^0_{13}\right)\frac{V_3}{2} + \left(\tau^0_{24}-\tau^0_{14}\right)\frac{V_4}{2}\right].
\label{Ic}
\end{eqnarray}
where $I^c=I^{\uparrow}_i+I^{\downarrow}_i$.
The dimensionless integer $N$ is the number of propagating wave modes in the leads, which is proportional to both the lead width ($W$) and the Fermi vector ($k_F$) through the equation $N = k_F W/\pi$, while $V_{3,4}$ are the potentials of the vertical leads.

Ref. \onlinecite{PhysRevB.93.115120} developed an analytic calculation of the spin Hall current average Eq. (\ref{Is}) and its fluctuations for a graphene device at the CNP. However, this was only possible by applying the diagrammatic method \cite{PhysRevB.88.245133,doi:10.1063/1.5010973} of the random matrix theory \cite{RevModPhys.69.731}. In this case, the random scattering matrix Eq. (\ref{M}) is  described by the chiral circular symplectic ensemble; that is, class CII in Cartan’s nomenclature \cite{PhysRevB.86.155118}. This means that the graphene device has a strong SOC, particle-hole and sublattices/mirror symmetries. { Assuming uniform SOI, we can obtain the spin Hall current average \cite{PhysRevB.93.115120} from Eq. (\ref{Is})
\begin{equation}
\langle{I^s}\rangle = 0. \label{Is1}
\end{equation}
Eq. \ref{Is1} is in agreement with the result of Ref. \onlinecite{PhysRevB.71.245327}, which has shown that the spin Hall conductivity is null for the uniform Rashba coupling, $\sigma_{sH}=0$. Meanwhile, Eq. \ref{Is1} is not in contradiction to the result of Refs. \onlinecite{PhysRevB.82.121310, https://doi.org/10.1002/andp.201100253}, in which the authors have shown for two different models that the spin Hall conductivity is proportional to the universal constant multiplied by momentum relaxation time and is inversely proportional to the spin relaxation time in the presence of disorder SOI, $\sigma_{sH}=e/4\pi \times \tau/\tau_s$.} 

Although the spin Hall current average is null, its fluctuation can be significant because of disorder. From Eq. (\ref{Is}), we obtain that
\begin{eqnarray}
\langle{{{\delta}{I^s}^2}}\rangle = \left(\frac{e^2V}{h}\right)^2\left[\frac{1}{32}+\mathcal{O}(N^{-1})\right],\label{Is2}
\end{eqnarray}
for a sufficiently large thickness $N$. From Eq. (\ref{Is2}), we conclude that the spin Hall current deviation is 
\begin{eqnarray}
\textbf{rms}[I^s] = \frac{e^2V}{h} 0.18,\label{UF}
\end{eqnarray}
which means that the spin Hall current fluctuations of graphene are universals. The longitudinal charge current average of a graphene device in the diffusive regime is appropriately described as \cite{PhysRevLett.108.166606}
\begin{eqnarray}
\langle{{{I^c}}}\rangle = \frac{e^2V}{h}\sigma,\label{Ic2}
\end{eqnarray}
in the function of dimensionless longitudinal conductivity $\sigma =N l_e/L$.

\subsection{Spin Hall angle fluctuations}

SHA is defined as the ratio between a transversal spin Hall current and a longitudinal charge current
\begin{eqnarray}
\Theta_{sH}=\frac{I^{s}}{I^c}.\label{I/I}
\label{angle}
\end{eqnarray}
We must implement the central limit theorem to develop the ensemble average on Eq. (\ref{I/I}). Therefore, by taking a sufficiently large thickness $N\gg1$, the Eq. (\ref{angle}) can be expanded as
\begin{equation}
\left\langle\Theta_{sH}\right\rangle = \frac{\langle{I^s}\rangle}
{\langle{I^c}\rangle}+
\frac{\langle{{\delta}I^s}\rangle{\langle I^c\rangle}-
\langle{{\delta}I^c}\rangle{\langle I^s}\rangle}
{\langle{I^c}\rangle^2}+\mathcal{O}(N^{-1}).\label{exp}
\end{equation}
From Eq. (\ref{Is1}), we know that the spin Hall current average is null, $\langle{I^s}\rangle = \langle{{\delta}I^s}\rangle = 0$, which leads us to conclude that
\begin{equation}
\left\langle\Theta_{sH}\right\rangle = 0.\label{TM}
\end{equation}
The graphene device that is under study is disordered, which induces fluctuations in the spin Hall current and the longitudinal charge current \cite{PhysRevLett.101.016804,ChoeChang,PhysRevB.102.115105}. Therefore, it is reasonable that the SHA fluctuates. In the usual way, we define the SHA deviation as
$$
\textbf{rms}[\Theta_{sH}]=\sqrt{\left\langle\Theta_{sH}^2\right\rangle-\left\langle\Theta_{sH}\right\rangle^2} = \sqrt{\left\langle\Theta_{sH}^2\right\rangle},\nonumber
$$
We follow the same methodology for the ensemble average and obtain
\begin{eqnarray}
\left\langle\Theta_{sH}^2\right\rangle &=& \frac{\langle{I^s}\rangle^2}
{\langle{I^c}\rangle^2}+2\frac{\langle{{\delta}I^s}\rangle\langle{I^s}\rangle{\langle{I^c}\rangle}-
\langle{{\delta}I^c}\rangle\langle{I^s}\rangle^2}
{\langle{I^c}\rangle^3}\nonumber\\&+&
\frac{\langle{{\delta}{I^s}^2}\rangle\langle{I^c}\rangle^2 +\langle{{\delta}{I^c}^2}\rangle{\langle{I^s}\rangle}^2-2\langle{{\delta}I^s{\delta}I^c}\rangle{\langle{I^s}\rangle}\langle{I^c}\rangle}
{\langle{I^c}\rangle^4}\nonumber\\
&+&\mathcal{O}(N^{-3}).\label{thetas}
\end{eqnarray}
Because the spin Hall current average is null, Eq. (\ref{thetas}) simplifies to 
\begin{eqnarray}
\textbf{rms}[\Theta_{sH}] &=& \sqrt{\frac{\langle{{{\delta}{I^s}^2}}\rangle}
{\langle{I^c}\rangle^2}}. \label{Tm}
\end{eqnarray}
From Eq. (\ref{Tm}) we can infer the SHA deviation with the knowledge of the spin Hall current fluctuations and the charge current average. Eq. (\ref{Tm}) is general, which means that it can be applied in any type of disorder device.

By substituting Eqs.(\ref{Is2}) and (\ref{Ic2}) in Eq.(\ref{Tm}), we obtain
\begin{eqnarray}
\textbf{rms}[\Theta_{sH}] = \frac{0.18}{\sigma} = \frac{0.18}{Nl_e/L} \label{rmsT}.
\end{eqnarray}
Eq.(\ref{rmsT}) indicates that the SHA attains a maximum deviation. Therefore, we can write it in a more general form, as follows
\begin{eqnarray}
\Theta_{sH} \times \sigma = 0.18. \label{Tc}
\end{eqnarray}
Eq. (\ref{Tc}) is valid for a graphene device at the CNP. However, it is the same result obtained by Ref. \onlinecite{PhysRevB.102.041107} for metals. 
Although the mechanisms of charge-to-spin current conversion in graphene and metals are different, in the presence of strong SOI the random scattering matrices of both devices (\ref{M}) are distributed by the circular symplectic ensemble of random matrix theory. Therefore, Eq. (\ref{Tc}) is a universal relation (i.e., independent of the microscopic features of the material).

\begin{figure}
\centering
\includegraphics[scale =0.6]{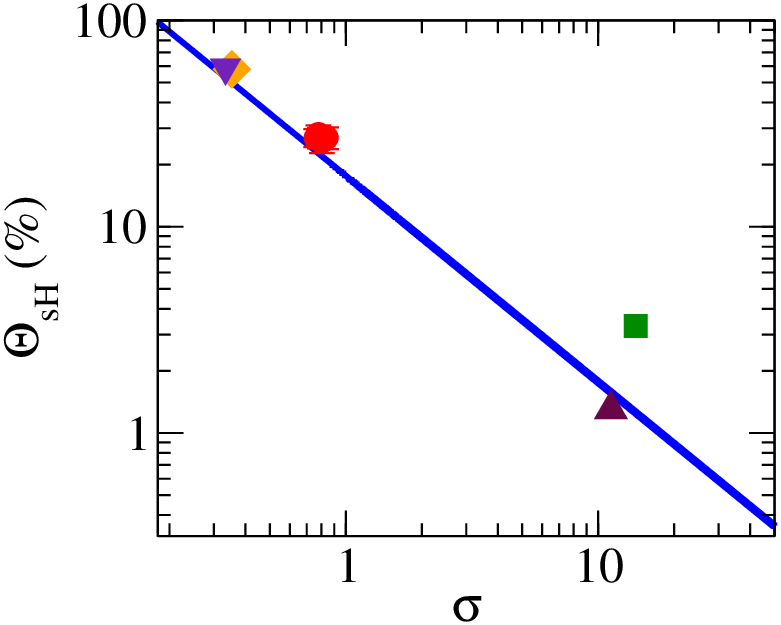}
\caption{SHA $\Theta_{sH} (\%)$ as a function of dimensionless conductivity $\sigma$. The diamond symbol (orange) denotes the experimental data from Ref. \onlinecite{10.1038/nphys2576}, the circle symbols (red) denote experimental data from  Ref. \onlinecite{10.1038/ncomms5748}, the  triangle up symbol (purple) denotes experimental data from Ref. \onlinecite{PhysRevResearch.2.013286}, the triangle down symbol (maroon) denotes experimental data  from Ref. \onlinecite{PhysRevResearch.2.033204} and the squared symbol (green) denotes experimental data  from Ref. \onlinecite{doi:10.1021/acs.nanolett.8b04368}. The continuum line (blue) is the analytical result of Eq.(\ref{Tc}).}\label{expprb}
\end{figure}

{\section{Comparison between theoretical and experimental results}}

Fig.(\ref{expprb}) shows $\Theta_{sH} (\%)$ as a function of dimensionless conductivity $\sigma$. The diamond symbol (orange) denotes experimental data of SHA at the CNP of graphene, which was obtained from Ref. \onlinecite{10.1038/nphys2576}. The dimensionless conductivity axis of the experiment was normalised as $\sigma=\sigma_{exp}(\Omega^{-1}\cdot $ cm$^{-1})/10^{4}(\Omega^{-1}\cdot $ cm$^{-1})$. 

The circle symbols (red) of Fig.(\ref{expprb})  are obtained from Figs. (12) and (13) between $E_F=[-50,\dots,50]$ meV in the Supplemental Material of Ref. \onlinecite{10.1038/ncomms5748} for copper-chemical vapour deposition graphene samples at room temperature.

The triangle up symbol (purple) of Fig.(\ref{expprb}) denotes experimental data obtained from Ref. \onlinecite{PhysRevResearch.2.013286} for Weyl semimetal WTe$_2$ with graphene. The triangle down symbol (maroon) denotes experimental data obtained from Ref. \onlinecite{PhysRevResearch.2.033204} for a hybrid device of TaTe$_2$ in a van der Waals heterostructure with graphene. The squared symbol (green) denotes experimental data obtained from Ref. \onlinecite{doi:10.1021/acs.nanolett.8b04368} for Graphene/MoS$_2$ van der Waals heterostructures.

We plot  Eq.(\ref{Tc}) as a continuum line (blue) together experimental data in Fig.(\ref{expprb}). From this figure, we can conclude the compatibility between the five experiments \cite{10.1038/nphys2576,10.1038/ncomms5748,PhysRevResearch.2.013286,PhysRevResearch.2.033204,doi:10.1021/acs.nanolett.8b04368} and the universal relation Eq.(\ref{Tc}).

\section{ Numerical results}

To confirm the predictions that Eq. (\ref{Tc}) is valid for a graphene device with disorder, we performed numerically accurate real-space simulations of the SHA. Fig. (\ref{sample}) illustrates the device design.
We expressed the Hamiltonian as $H=H_{g}+H_{R}+H_{d}$, where $H_{g}$ describes the usual nearest-neighbor hopping, $H_{R}$ is the nearest-neighbour hopping term describing the Bychkov-Rashba SOC, which
explicitly violates $\vec{z} \rightarrow {-}  \vec{z}$ symmetry, and $H_d$ is the on-site potential of carbon atoms in the hexagon hosting
adatoms, which simulates a charge modulation that is induced locally around the adatom. In terms of annihilation (creation) operators $c_{i,\sigma}$  ($c_{i,\sigma}^\dagger$) that remove (add) electrons to site $i$ with spin $\sigma = \uparrow,\downarrow$,  the terms $H_g$, $H_{R}$ and  $H_d$ read as follows \cite{PhysRevLett.119.246801,PhysRevB.93.155104,PhysRevB.98.045407,PhysRevB.103.L081111}
\begin{eqnarray}
 H_g&=&-\sum_{\langle i,j \rangle,\sigma} t \, c_{i,\sigma}^{\dagger} c_{j,\sigma}\,, 
\\
H_{R}&=& - \sum_{\langle i,j \rangle, \sigma, \sigma^\prime } \imath \lambda_{i,j} \, c_{i,\sigma}^{\dagger}\left(\left[\mathbf{s}\right]_{\sigma \sigma^\prime}\times \hat \mathbf{r}_{ij}\right)_z c_{j,\sigma^\prime}\,,\label{TBHCS}
\\
H_d&=&-\sum_{\langle i \rangle,\sigma} \epsilon_{i} \, c_{i,\sigma}^{\dagger} c_{i,\sigma}\,, 
\label{TBHC}
\end{eqnarray}
where the indices $i$ and $j$ run over all lattice sites,\,$ \langle \cdots \rangle$ denotes a sum over nearest-neighbor sites, $\hat \mathbf{r}_{ij}$ is the unit vector along the line segment connecting the sites $i$ and $j$, and $t=2.8$ eV is the hopping integral. Therefore, $\lambda_{i,j}$ is the Bychkov-Rashba coupling strength between sites $i$ and $j$. The disorder is an electrostatic potential $\epsilon_i$ that varies randomly from site to site according to a uniform distribution in the interval $\left(-U/2,U/2\right)$, where $U$ is the disorder strength. The Fermi energy, disorder strength $U$ and Bychkov-Rashba SOC $\lambda_{i,j}$ values are given in unities of $t$, while the width and length of the device $W=L=40$ are given in unities of the graphene lattice constant, $a_0 = 2.49  \AA$. All of the numerical results were obtained from 15,000 disorder realisations. The numerical calculations were implemented in KWANT software \cite{Groth_2014}.

\begin{figure}
\centering
\includegraphics[scale = 0.35]{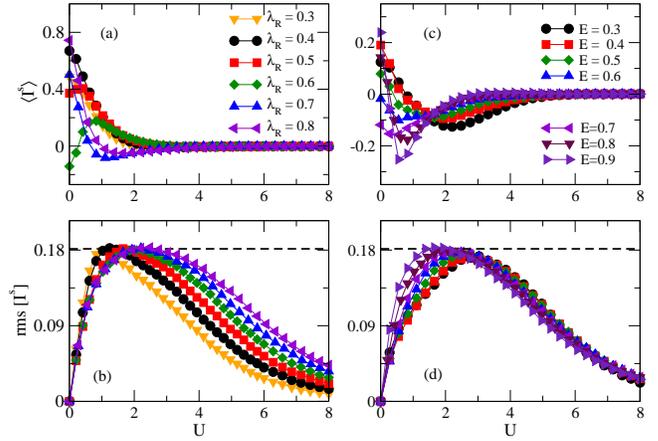}
\caption{Figures (a,c) show the spin current average, while Figures (b,d) show the spin current deviation as a function of the disorder $U$. Figures (a,b) show different values of SOC $\lambda$ at fix $E=0.8$, while Figures (c,d) show different values of $E$ at fix $\lambda=0.6$. In both cases, the spin Hall current deviation results in $\textbf{rms}[I^s] = 0.18$ (dashed line).}\label{figIs}
\end{figure}

We have developed numerical calculations for two different configurations. The first configuration is for a graphene device with uniform Bychkov-Rashba SOC, where we keep fixed the SOC strength $\lambda_{i,j}=\lambda$, while the disorder is a random electrostatic potential (as discussed above). The second configuration is for a graphene device with a random Bychkov-Rashba SOC, where the SOC strength $\lambda_{i,j}$ is randomly accorded by a uniform distribution in the interval $\left(-\lambda/2,\lambda/2\right)$, while the disorder strength is kept null $U=0$.

\begin{figure}
\centering
\includegraphics[scale = 0.35]{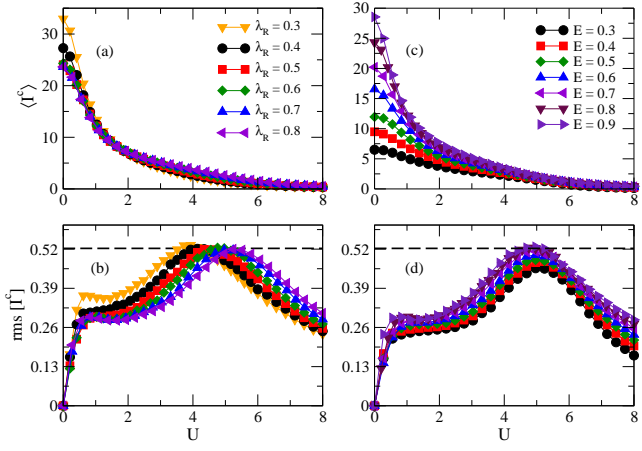}
\caption{Figures (a,c) show the charge current average, while Figures (b,d) show the charge current deviation in function of disorder $U$. Figures (a,b) are for different SOI values $\lambda$ at fix $E=0.8$, while Figures (c,d) are for different values of $E$ at fix $\lambda=0.6$. In both cases, the charge current deviation holds a maximum in $\textbf{rms}[I^{c}] = 0.52$ (dashed line). }\label{figIc}
\end{figure}

\subsection{Uniform  Bychkov-Rashba SOC}

In this section, we study the SHE with uniform Bychkov-Rashba SOC. The SOC strength of Eq. (\ref{TBHCS}) is fixed $\lambda_{i,j}=\lambda$ and the disorder of the graphene device is introduced by an electrostatic potential $\epsilon_i$, which varies randomly from site to site according to a uniform distribution in the interval $\left(-U/2,U/2\right)$, where $U$ is the disorder strength.  

We begin with Fig. (\ref{figIs}), which shows numerical calculations of the spin Hall current as a function of disorder $U$ from Eq.(\ref{Is}). Figs.(\ref{figIs}.a,c) show the spin Hall current average, while Figs.(\ref{figIs}.b,d) show its deviation for different values of $\lambda$ and Fermi energy.
In the former, we observe oscillations in the tails of the spin Hall current average, which are similar to those observed in the numeric results for the diffusive device in Ref. \onlinecite{PhysRevB.102.041107}. The underlying mechanism of the oscillations are the fluctuations in potentials $V_{3,4}$. In  Figs.(\ref{figIs}.a,b), the energy was fixed in $E=0.8$ for different SOI values $\lambda$. For all values of $\lambda$, the spin Hall current deviation holds the universal maximum deviations Eq. (\ref{UF}) (dashed line), Fig. (\ref{figIs}.b). Furthermore, in Figs.(\ref{figIs}.c,d), the SOI value was fixed in $\lambda = 0.6$ for different energy values. The spin Hall current has its universal maximum deviation for all energies; see Fig. (\ref{figIs}.d).

\begin{figure}
\centering
\includegraphics[scale = 0.35]{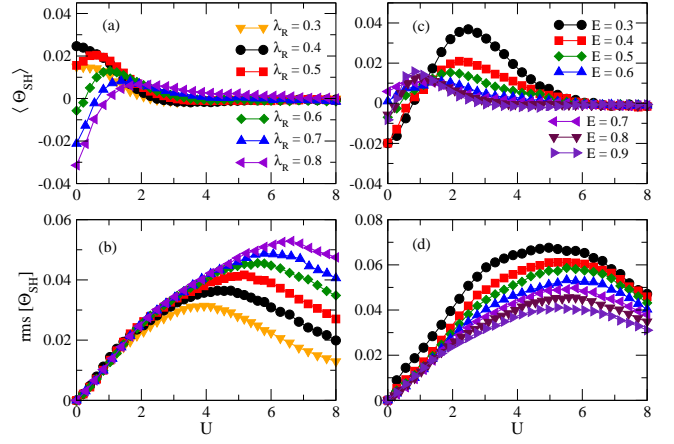}
\caption{Figures (a,c) show the SHA average, while Figures (b,d) show the one deviation in function of disorder $U$. In Figures (a,b), each curve is for a different value of SOI $\lambda$ at fixed energy $E=0.8$.  In Figures (c,d), each curve is for a different value of $E$ at fixed $\lambda=0.8$. }\label{figASH}
\end{figure}

As developed earlier for the spin Hall current, we numerically calculate  the longitudinal charge current, Eq.(\ref{Ic}), which is depicted in Fig.(\ref{figIc}). Figs.(\ref{figIc}.a,c) show the charge current average as a function of $U$ for different values of $\lambda$ and energy, respectively, while  Figs.(\ref{figIc}.b,d) are their respective deviations. Although the spin Hall current average shows oscillations in the tails, as depicted in Figs.(\ref{figIs}.a,c), the charge current average does not present them, as depicted in Figs.(\ref{figIc}.a,c). Furthermore, the charge current maximum deviation in Fig.(\ref{figIc}.b) occurs for disorder strength values ($U\geq 4$) that are larger than spin Hall maximum deviation ($U \approx 2$), see Fig.(\ref{figIs}.b). From the numeric data of Figs.(\ref{figIc}.b,d), we estimate the charge current maximum deviation as $\textbf{rms}[I^c] = e^2V/h \times 0.52$ (dashed line).

We are now ready to analyse the SHA, Eq.(\ref{angle}), which is depicted in Fig.(\ref{figASH}). Figs.(\ref{figASH}.a,c) show the SHA average as a function of $U$ for different values of $\lambda$ and energy, respectively, while Figs.(\ref{figASH}.b,d) are their respective deviations. As we can see in Figs.(\ref{figASH}.a,c), the SHA average  keeps the oscillations present in the spin Hall current average. However, the SHA maximum deviations happen only for $U\geq 4$, see Fig.(\ref{figASH}.b). This means that the efficiency increase is not related to an increase in the spin Hall current fluctuations but is related to an increase in the charge current fluctuations. The charge-to-spin current conversion is more efficient when the charge current fluctuates more, which is in accordance with Eq.(\ref{rmsT}).

\begin{figure}
\centering
\includegraphics[scale = 0.45]{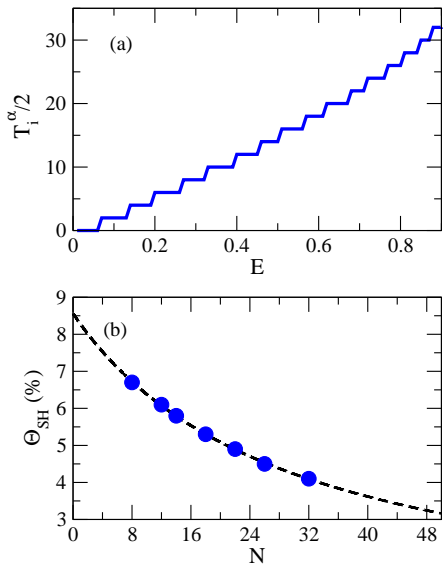}
\caption{(a) The transmission coefficient $T_i^{\alpha}(E)/2=N$ as a function of energy. (b) The SHA maximum deviations of Fig.(\ref{figASH}.d) as a function of thickness $N$. The dashed line is the numeric data fit.}\label{figAlNL}
\end{figure}

Although Figs. (\ref{figIs}.d) and (\ref{figIc}.d) show an increase in the maximum deviation of spin and charge currents as a function of energy with converging to a finite value, the maximum SHA deviation decreases as a function of energy without converging; as demonstrated in  Fig.(\ref{figASH}.d). For example, the SHA has its maximum deviation $\Theta_{sH} \approx 7\% $ when the energy is $E=0.3$. This means that the SHA decreases as the energy increases, which is in agreement with Eq. (\ref{rmsT}).

Finally, we are now in a position to connect the numerical results with Eq. (\ref{rmsT}). Fig. (\ref{figAlNL}.a) shows the transmission coefficient $T_i^{\alpha}(E)=\sum_{j}\tau_{ij}^{\alpha}(E)=2N$ as a function of energy, which gives the relationship between  $E=0.3, 0.4,0.5,0.6,0.7,0.8,0.9$ and  $N=8,12,14,18,22,26,32$. Fig. (\ref{figAlNL}.b) shows the maximum SHA deviation from Fig. (\ref{figASH}.d) as a function of $N$. The dashed line is the numerical data fit, $$\Theta_{sH} = \frac{1}{11.7+0.4 \times N}.$$ Taking the limit of large thickness $N$ for which  Eq. (\ref{rmsT}) is valid, it goes to  $$\Theta_{sH} \approx \frac{1}{0.4 \times N} = \frac{0.18} {0.072 \times N}.$$ By comparing the latter with the Eq.(\ref{Tc}), we obtain $\sigma=Nl_e/L= 0.072 \times N$, which drives to a universal relation $$\Theta_{sH} \times \sigma = 0.18,$$ (as previously stated). Furthermore, we can estimate the mean free electron path of a graphene device as $l_e= 0.072 \times L=2.88$.

\begin{figure}
\centering
\includegraphics[scale = 0.085]{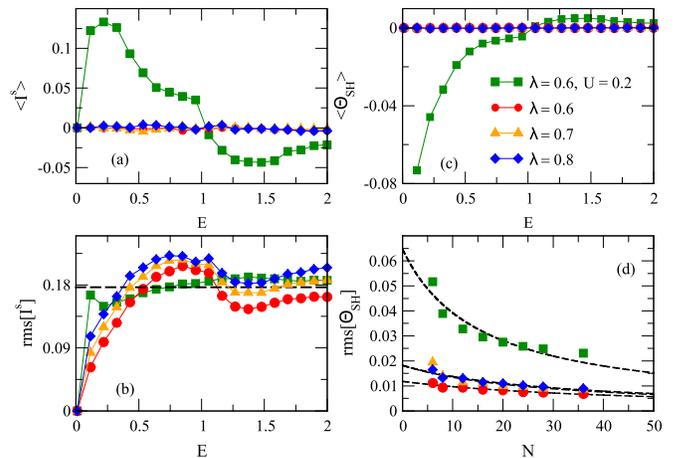}
\caption{The top figures show the  (a) spin Hall current average and (c) spin Hall angle average in function of energy. The square symbols denote a uniform Rashba SOC $\lambda = 0.6$ and $U=2$, while the other symbols denote random SOC with $\lambda = 0.6$, 0.7 and 0.8 and without electrostatic disorder $U=0$. Figure (b) shows the spin Hall current deviation as a function of energy. In all cases, the deviation results converge to $\textbf{rms}[I^s] = 0.18$ (dashed line). Figure (d) shows the spin Hall angle deviation as a function of thickness $N$. The dashed lines are numeric data fitted by $\textbf{rms}[\Theta_{sH}]\sim N^{-1}$.}\label{figrandom}
\end{figure}

\subsection{Random Bychkov-Rashba SOC}

Motivated by Refs. \onlinecite{PhysRevB.77.193302,PhysRevB.82.121310,https://doi.org/10.1002/andp.201100253,PhysRevB.97.245307,SEIBOLD201763,KUMARSHARMA2021167711,PhysRevB.83.085306}, we analysed the SHE with random Bychkov-Rashba SOC. In this case, the SOC parameter $\lambda_{i,j}$ of Eq. (\ref{TBHCS}) is responsible for graphene disorder and is distributed randomly according to a uniform distribution in the interval $\left(-\lambda/2,\lambda/2\right)$, where $\lambda$ is the SOC strength. {Hence, we keep the amplitude of the random electrostatic potential null in the numeric calculation, $U = 0$ in Eq. (\ref{TBHC}).}  

{The random Bychkov-Rashba SOC Hamiltonian (\ref{TBHCS}) is similar to the one that was introduced by Ref. \onlinecite{PhysRevB.83.085306} in the small correlation length limit. In this limit, the Dyakonov-Perel' mechanism dominates the spin fluctuating precession in the graphene device; that is, the electron spin interacts with the local rather than with the rapidly changing random spin-orbit field \cite{PhysRevB.83.085306}.}

Figs. (\ref{figrandom}.a) and (\ref{figrandom}.c) show the numeric calculation data for the average of the spin Hall current and SHA as a function of energy, respectively. For a direct comparison between uniform and random SOC,  we plotted a uniform SOC numeric data, with $\lambda = 0.6$ and $U=2$ (square symbols), and three different numeric data for random SOC, with SOC strengths $\lambda = 0.6$, 0.7 and 0.8. The former has a no null
spin Hall current average, while the latter has a null average for all energy ranges, see Fig. (\ref{figrandom}.a). Similar behavior is found for the SHA average, see Fig. (\ref{figrandom}.c). However, we are interested in the spin Hall current and SHA deviations.  

Fig. (\ref{figrandom}.b) shows the spin Hall current deviation as an energy function. The numerical data of uniform and random SOC go to 0.18 as the energy increases, which is in agreement with Eq. (\ref{UF}). Therefore, the universal spin Hall current fluctuations are significant for both cases. This indicates that {vanishing} SHE is irrelevant for realistic finite-size systems, where self-averaging over an infinite system size is avoided. 

Finally, we plotted the maximum SHA deviation as a function of thickness $N$ in Fig. (\ref{figrandom}.d). From this figure it can be seen that for uniform and random SOC, maximum SHA deviation decreases with thickness increases $N$, which is in agreement with our analytical result Eq. (\ref{rmsT}). Therefore, the numeric data can be well fitted by relation $\textbf{rms}[\Theta_{sH}] \sim N^{-1}$. 

\section{ Conclusions}

This work studied the SHA fluctuations of single-layer graphene with disorder and strong SOC. We analytically show that the product between maximum SHA deviation and dimensionless conductivity of a graphene device follows the same universal relation of metals, see Eq. (\ref{Tc}). 
To confirm this analytical result, we developed numerically exact real-space simulations from the tight-binding model with Bychkov-Rashba SOC. The numeric data are consistent with Eq. (\ref{Tc}), which confirms its validity to Dirac materials as graphene.

We also developed numerical simulations of SHE in the presence of uniform and random Bychkov-Rashba SOC.
We found that the averages of spin Hall current and SHA as an energy function have different behaviors for uniform and random SOC---the former has a finite value for the spin Hall current and SHA averages, while the latter has a null average; see Figs. (\ref{figrandom}.a) and (\ref{figrandom}.b).

Although the averages of spin Hall current and SHA have different behaviors for uniform and random SOC, their deviations are large and equivalent; as shown in Figs. (\ref{figrandom}.b) and  (\ref{figrandom}.d). This confirms that vanishing SHE is irrelevant for a realistic finite-size graphene device where self-averaging over an infinite system size is avoided. Therefore, the maximum SHA deviation of the graphene devices with uniform or random SOC follows Eq. (\ref{Tc}).

Finally, we confronted five different experimental data \cite{10.1038/nphys2576,10.1038/ncomms5748,PhysRevResearch.2.013286, PhysRevResearch.2.033204,doi:10.1021/acs.nanolett.8b04368} with Eq. (\ref{Tc}). As shown in Fig. (\ref{expprb}), both the analytical results and the experimental data agree satisfactorily. Therefore, we believe that our results can contribute to a more profound understanding of SHE and SHA fluctuation in Dirac materials.

\section*{Acknowledgments}
This work was supported by CNPq (Conselho Nacional de Desenvolvimento Cient\'{\i}fico e Tecnol\'ogico) and FACEPE (Funda\c{c}\~ao de Amparo \`a Ci\^encia e Tecnologia do Estado de Pernambuco).

\section*{Data Availability}
The data that support the findings of this study are available from the corresponding author upon reasonable request.

\section*{References}

\bibliography{ref}

\end{document}